\documentclass[12pt]{article}     
\usepackage{amssymb,epsf}

\newfont{\isf}{cmssi12 scaled 1000 \baselineskip12pt}  

\newfont{\bsf}{cmssbx10 scaled 1200 \baselineskip12pt} 

\newfont{\lsf}{cmssbx10 scaled 1440 \baselineskip14pt} 

\begin{document}
\begin{flushright}
YITP-99-41\\
June~~ 1999
\end{flushright}
\begin{center}
\lsf 
THEORETICAL PREDICTIONS OF RESIDUES CROSS SECTIONS OF SUPERHEAVY ELEMENTS
{\def\thefootnote{\fnsymbol{footnote}}
\footnote[3]{\it Invited talk given at Nuclear Shells-50 years, Dubna, April 
         21-24, 1999}}

\vspace{3mm}
\isf
Y. Abe$^1$, K. Okazaki$^2$, Y. Aritomo$^3$, T. Tokuda$^2$, T. Wada$^2$ and M. Ohta$^2$

\sf
$^1$Yukawa Institute for Theoretical Physics, Kyoto Univ., Kyoto
606-0l, Japan\\
$^2$Department of Physics, Konan Univ., Kobe 658, Japan\\
$^3$Flerov Laboratory of Nuclear Reactions, JINR, Dubna 141980, Russia\\
\end{center}

\sf
Dynamical reaction theories are reviewed for synthesis of superheavy
elements. Characteristic features of formation and surviving are
discussed with reference to possible incident channels.  Theoretical
predictions are presented on favorable incident channels and on
optimum energies for synthesis of Z = 114.

\vspace{5mm}
{\bf 1. Introduction}\\
Superheavy elements around Z = 114(or 126) and N = 184 have been
believed to exist according to theoretical predictions of stability
given by the shell correction energy in addition to average nuclear
binding energy$^{1)}$.  This means that heavy atomic nuclei with fissility
parameter {\isf x} $\gtrsim$ 1 could be stabilized against fission by a
huge barrier which is resulted in by the additional binding of the
shell correction energy around the spherical shape.  In other words, if
superheavy compound nuclei(C.N.) are formed in such high excitation
that the closed shell structure is mostly destroyed, they have no barrier 
against fission and thus are inferred to decay very quickly,
though time scales of fission are now believed to be much longer than 
that of Bohr-Wheeler formula due to a strong friction for the
collective motion$^{2)}$.
Therefore, the point is how to reach the ground state of the
superheavy nuclei, or how to make a soft-landing at them. In order to
minimize fission decays of C.N. or maximize their
survival probabilities, so-called cold fusion reactions have been
used, which succeeded in synthesizing SHEs up to Z = 112$^{3)}$.  They have
the merit of large survival probabilities, but suffer from the demerit 
of small formation probabilities because of the sub-barrier fusion.
On the other hand, so-called hot(warm) fusion reactions have the merit 
of expected large formation probabilities and the demerit of small
survival probabilities due to relatively high excitation of C.N. formed.  Anyway, an optimum condition for large residue cross
sections of SHEs is a balance or a compromise between formation and
survival probabilities as a function of incident energy or excitation 
energy of C.N. formed over possible combinations of projectiles and
targets$^{4)}$.

\vspace{5mm}
{\bf 2. Two Reaction Processes: Formation and Surviving (Decay)}\\
They are
not always independent, especially in so-called massive systems, but
for simplicity we briefly discuss them separately. Formation of
C.N. is by the fusion reaction.  Fig.1 reminds us of its
characteristic features, depending on the system.  In lighter systems, 
i.e., those with Z$_1$$\cdot$Z$_2$ $\lesssim$ 1,800,they undergo
fusion if they have enough energy to overcome the Coulomb barrier (say, Bass
barrier$^{5)}$, while in heavier systems, they have to overcome so-called
conditional saddle to get fused even after overcoming the Coulomb barrier.
Since the systems are under the action of strong nuclear interactions, 
their incident kinetic energies are quickly transformed into internal
motions, and thereby much more energy than the difference between the
barrier and the saddle point is required for formation of C.N., which
corresponds to the extra-push or extra-extra-push energy$^{6)}$.  One more
point to notice is that the potential energy surface for SHE has
almost no pocket schematically shown in Fig.1 if the C.N. formed are
in  rather
high excitation.  This would be the reason why a simple practical
formula is not available for SHE formation probability. A dynamical
framework had been called for so long untill the recent works appeared$^{4)}$.  It is also
worth to mention that Fig.1 is just a one-dimensional
schematization.  Real processes are in many dimensions including
mass asymmetry degree of freedom etc. in addition to the elongation or the separation
between two fragments.  An important case that we will discuss below is
that the incident channel is with Z$_1$$\cdot$Z$_2$ $\simeq$ 1,800 and 
the compound nucleus is with Z = 114.  The potential energy surface for
the compound nucleus has almost no minimum like that shown in Fig.1
due to excitation, while the Bass barrier is high and quite inner,
close to the point where the energy surface becomes flat.

\vspace{-10mm}
\begin{figure}[h]
   \epsfysize=50mm
   \epsfbox{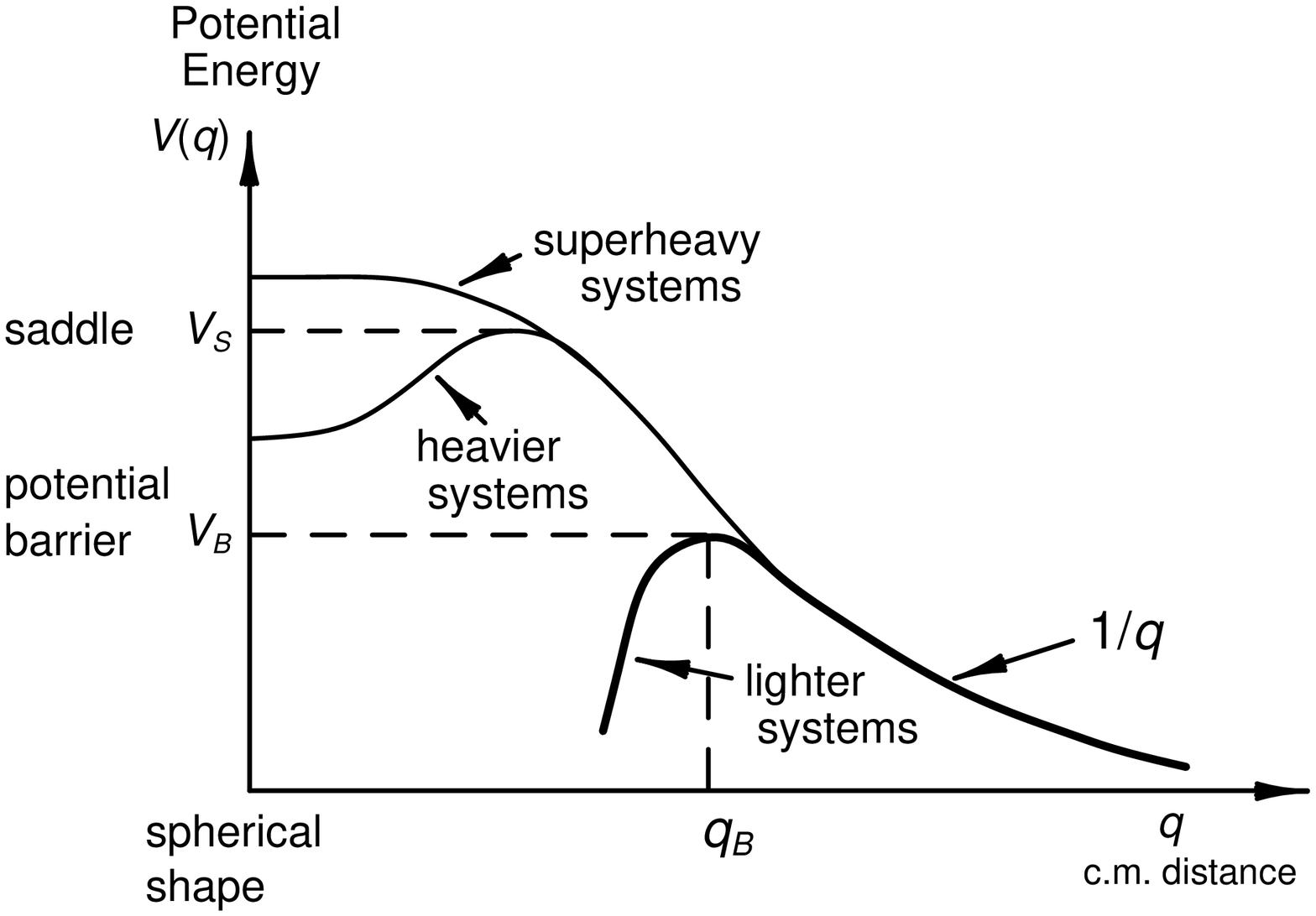}\hspace{30mm}
    \epsfysize=70mm
    \epsfbox{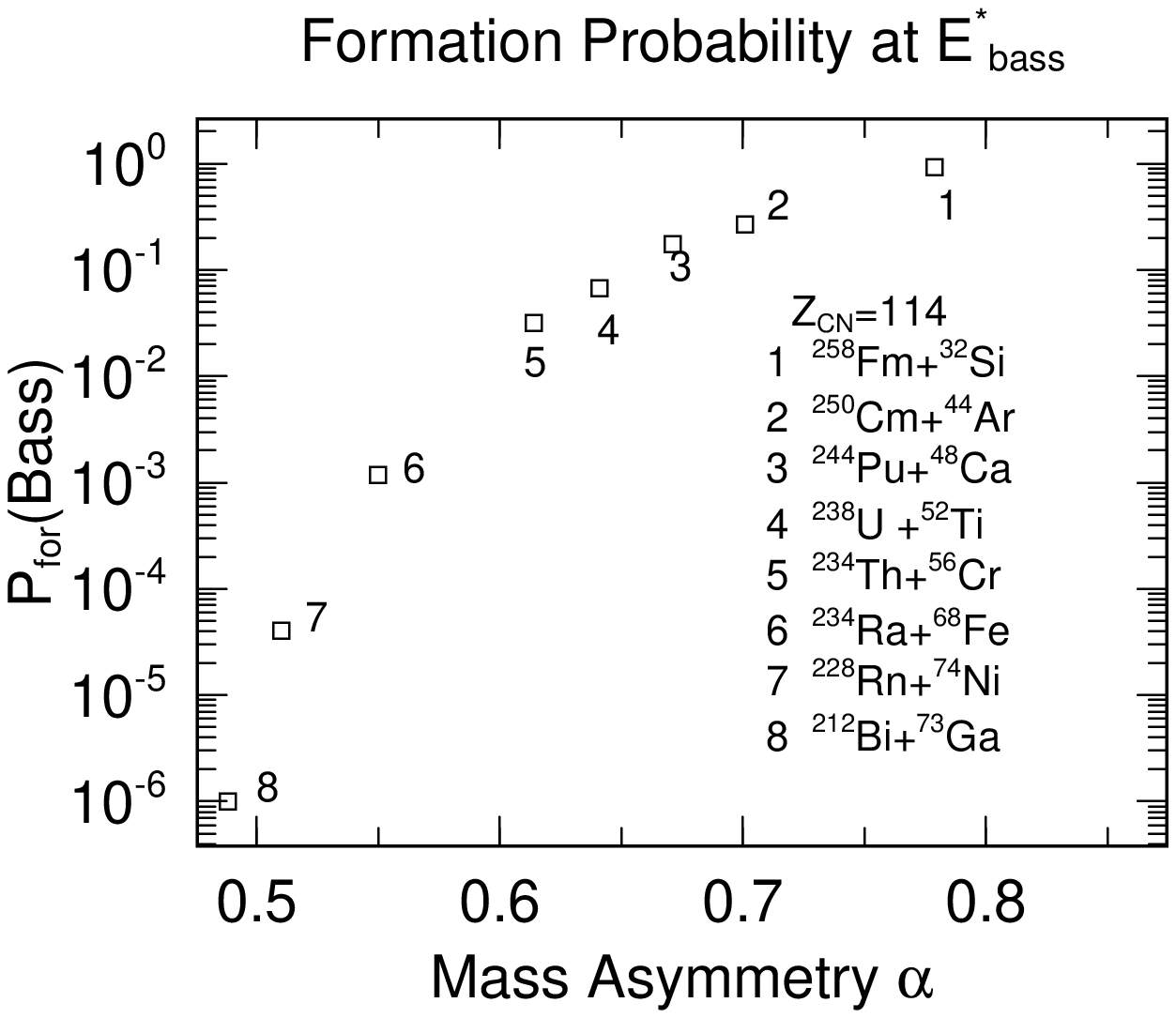}
\end{figure}

\hspace{40mm}{\bf Figure 1}
\hspace{55mm}{\bf Figure 2}

\vspace{5mm}

We have calculated formation probabilities in the following
way.$^{7)}$
If an incident energy is below the barrier, we take into account the
barrier penetration factor using WKB approximation.  The potential
(barrier) is calculated with the Coulomb and the nuclear proximity
potentials$^{8)}$ between the incident ions, where effects of
deformations etc. are not taken into account in order to see simply a
general trend.  After the incident ions reach the contact point,
evolutions of shapes of the total system are under the
dissipation-fluctuation dynamics, as mentioned above.  We have
employed a multi-dimensional Langevin equation to describe
trajectories in three-dimensional space where the distance (or
elongation) degree of freedom is taken into account as well as the
mass-asymmetry and the fragment deformation.
Some trajectories
go to the spherical shape of the compound nucleus and its around,
while some others to
reseparations after random walks in the space. Examples of calculated formation probabilities are shown in Fig.2 for Z = 114 C.N. with the
possible incident channels at the incident energies corresponding to
their Bass barriers.  We can readily see that the larger the mass asymmetry 
($\alpha$) is, the larger the formation probability (P$_{for}$) is.  This is 
just consistent to the feature in Z$_1$$\cdot$Z$_2$ dependence of
fusion reactions mentioned above. The smalll P$_{for}$'s in small mass- 
asymmetric cases correspond qualitatively to the ``heavier systems'' in
Fig.1, i.e. are due to the strong friction for the collective
motions. What is noticeable
here is the great increase of several orders of magnitude as a
function of $\alpha$.  This indicates that mass asymmetric incident
channels do not suffer much from the dissipation and are extremely
favorable in formation of C.N., but on the other hand, as shown in
Fig.3, C.N. formed with mass asymmetric channels have higher
excitation energies than those with less asymmetries, due to Q-values, 
which means that asymmetric channels are unfavorable for surviving.
In order to know more precisely about excitation-energy dependence of 
survival probability (P$_{sur}$) in
SHEs, we have to take into account effects of  cooling speeds which
are  essential for
SHEs, because superheavy C.N. can be stabilized only by the restoration of
the shell correction energy which is determined by the cooling, i.e.,
mainly by neutron evaporation. 
For particle evaporations we have used the statistical theory 
as usual. One more crucial factor in determining P$_{sur}$ is the time 
scale of  \\

\vspace{50mm}
\begin{figure}[h]
   \epsfxsize=70mm
   \epsfbox{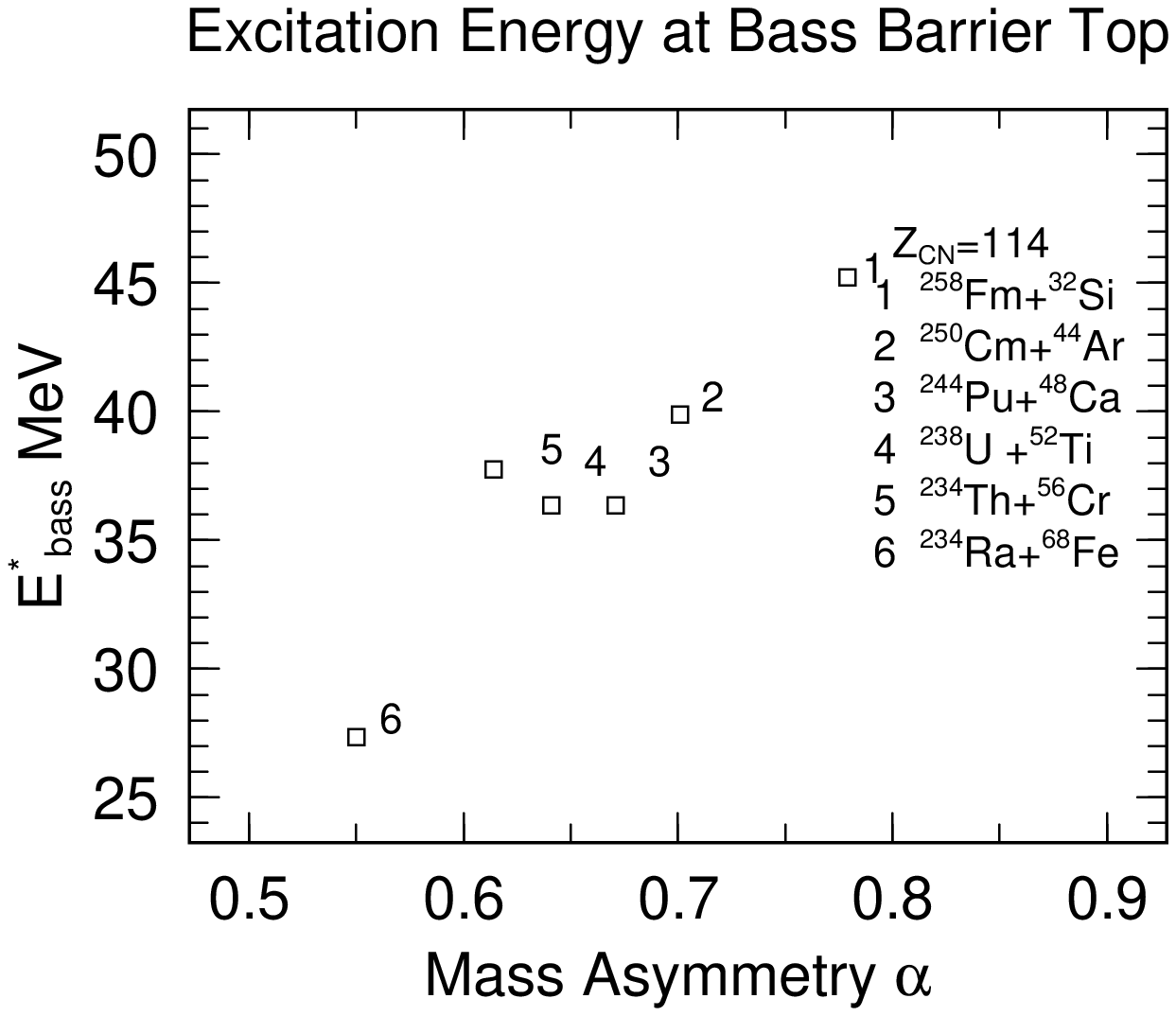}
\end{figure}
\vspace{-160mm}
\begin{figure}[h]
\hspace{100mm}
    \epsfxsize=80mm
    \epsfbox{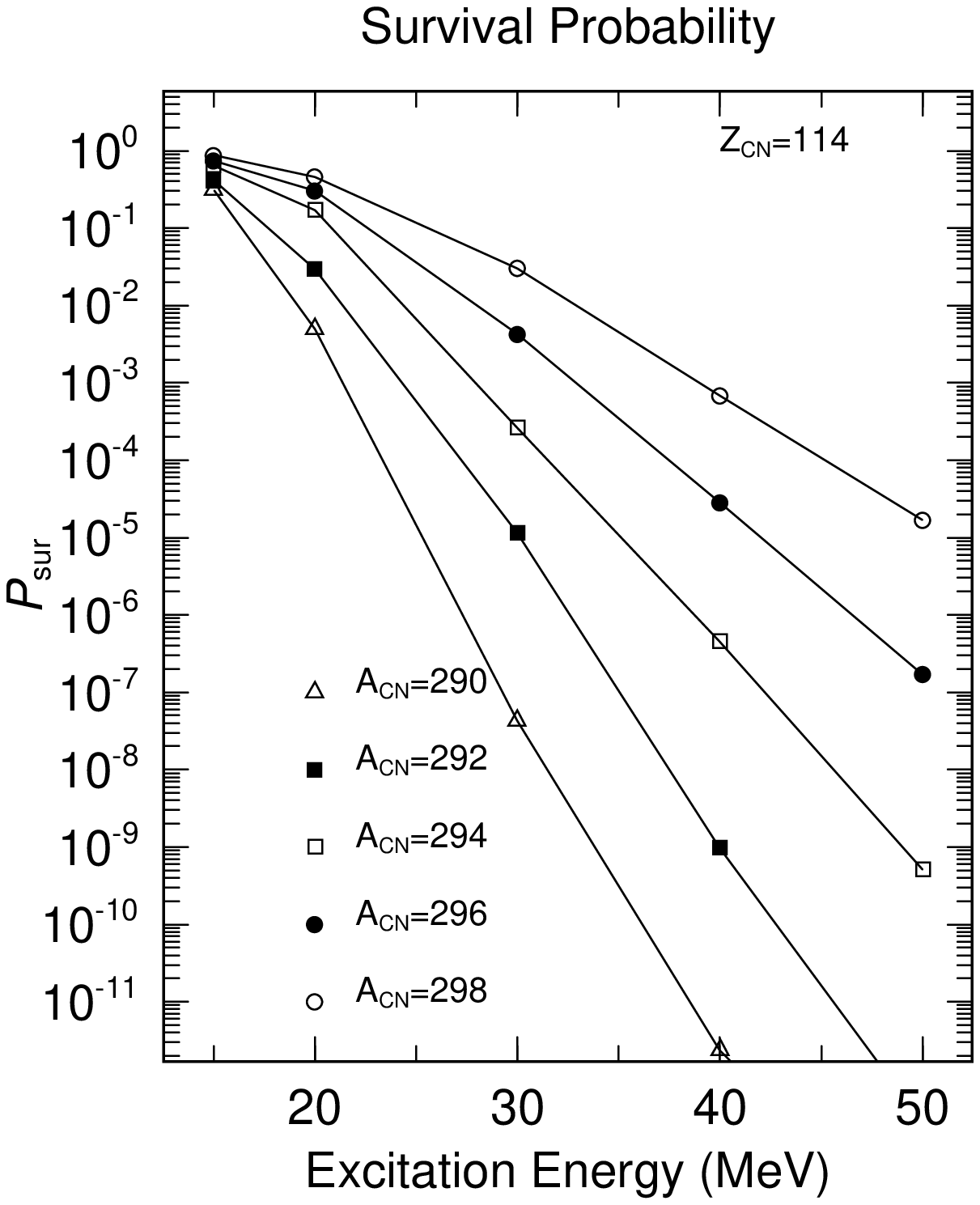}
\end{figure}

\vspace{6mm}
\hspace{35mm}{\bf Figure 3}
\hspace{60mm}{\bf Figure 4}

\vspace{5mm}
fission.  Since we know fission of excited nuclei is a
dynamical process under strong friction, we have employed
one-dimensional Smoluchowski equation for describing the evolution of
fissioning degree of freedom, which is known to be correct enough for 
the present purpose.$^{2)}$  Results of P$_{sur}$ for Z
= 114 are shown in Fig. 4 as a function of excitation energy(E$^*$)
over several mass numbers A.  It is surprising
that i) P$_{sur}$'s  decrease very quickly as E$^*$ increases and ii) mass
number dependence of the decrease is enormous.  This means that
C.N. with large mass numbers are favorable for surviving.  This is
essentially due to quick coolings in neutron-rich C.N. where the 
separation energy Bn's are small. Thus, unfavorable large E$^*$s could
be somehow compensated by the quick coolings if C.N. are of small
B$_n$, of course, with the aid of rather long time scales of fission.
On the other hand, if we initially form neutron-deficient isotopes,
cooling speeds are slow and thereby their survival probabilities drop
very rapidly as E$^*$ increases.  In such cases, we have to form C.N.
 in as low excitation as possible in order to obtain large residue cross
sections, which is qualitatively consistent with GSI
experiments.$^{3)}$

\vspace{5mm}
{\bf 3. Examples of the Calculated Cross Sections}\\
We have calculated excitation functions of evaporation residue cross
sections by combining the two reaction processes; formation and
surviving. 
Results for possible incident channels to form Z = 
114 isotopes are shown in Fig.5 as a function of E$^*$. The left-hand
side increases toward the peaks are due to formation probabilities, 
i.e., the barrier penetration and the dynamical evolution for fusion, 
while the right-hand
decreases due to E$^*$ dependence of survival probabilities. The
arrows with the numbers show the positions of the Bass barriers in the 
channels, respectively.  The incident  channels 
$^{250}$Cm + $^{44}$Ar and $^{244}$Pu + $^{48}$Ca are predicted
to have cross sections more than 1 pb 
which is thought to be a limit in measurements. The
importance of larger neutron numbers is readily understoood.  It is extremely
interesting that Dubna group has recently observed an event which could be
related to a synthesis of Z = 114 with the latter channel$^{9)}$.\\

\vspace{5mm}
\begin{figure}[h]
   \epsfysize= 13cm
   \epsfbox{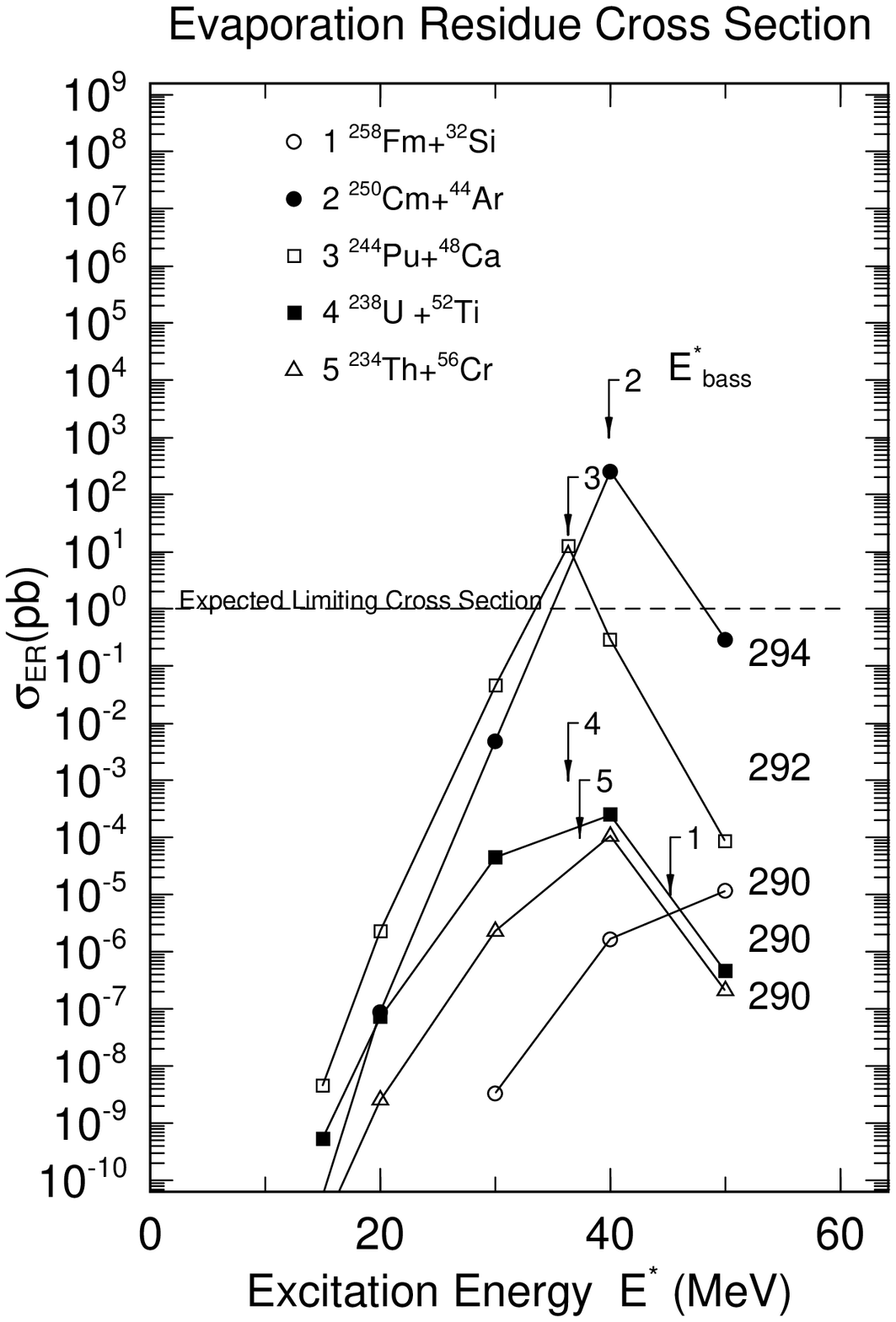}
\end{figure}
\hspace{40mm}{\bf Figure 5}

\vspace{-140mm}
\hangindent=95mm
\hangafter=-40
\vspace{-10mm} 
\noindent{\bf 4. Remarks}\\
It would be worth to mention again i) that the important point is a balance between 
formation and surviving and ii) that the neutron separation energy Bn's
which determine cooling speeds are another important quantities in
addition to the magnitudes of the shell correction energy. The
second point encourages us to explore exotic targets and projectiles
with more neutron excess. For more precise quantitative prediction of
residue cross sections,
one-dimensional WKB approximation for the
penetration factor should be improved so as to
accomodate effects of the deformations of the incident ions etc.
The last remark is on more mass-symmetric incident channels which are
not shown here.  They generally suffer more from the effects of
dissipation which unfavor fusion probabilities but on the other hand,
if neutron-rich C.N. could be formed, again there is a hope to obtain
rather large residue cross sections.$^{4)}$

\vspace{24mm}
{\bf References:}\\
$[1]$ P. M\"oller et al., At. and Nucl. Data Table {\bsf 59} (1996) 185. 
\\
\null\hspace{5mm}
S. Cwiok, et al., Nucl. Phys. {\bsf A611} (1996) 
211. \\
$[2]$ T. Wada, Y. Abe and N. Carjan, Phys. Rev. Lett. {\bsf 70} (1993) 
3538. \\
\null\hspace{5mm}
Y. Abe et al., Phys. Reports {\bsf C275}(1996) 49. \\
$[3]$ S. Hofmann et al., Z. Phys. {\bsf A354} (1996) 229. \\
$[4]$ Y. Abe et al., J. Phys. {\bsf G23} (1997) 1275. \\
\null\hspace{5mm}
Y. Aritomo et al., Phys. Rev. {\bsf C55} (1997) R1011, and {\isf ibid}
{\bsf C59}(1999) 796.\\
$[5]$ R. Bass, Nuclear Reactions with Heavy Ions (Springer, 1980).\\
$[6]$ W.J. Swiatecki, Nucl. Phys. {\bsf A376} (1982) 275. \\
\null\hspace{5mm}
S.Bjornholm and W.J. Swiatecki, Nucl. Phys. {\bsf A391} (1982) 471. \\
$[7]$ T. Wada et al., Proc. DANF98, Slovakia, Oct. 1998.\\
\null\hspace{5mm}
K. Okazaki, et al., publication under preparation. \\
$[8]$ J. Blocki et al., Ann. Phys.(N.Y.) {\bsf 105} (1997) 427. \\
$[9]$ Yu. Oganessian et al., preprint JINR, E7-99-53. \\
\end{document}